\title{Hidden Markov Models for Pipeline Damage Detection Using Piezoelectric Transducers}
\author{
	\textsc{Mingchi Zhang}
	\qquad
	\textsc{Xuemin Chen}\thanks{Contact author, xuemin.chen@tsu.edu}
	\qquad
	\textsc{Wei Li}
	\mbox{}\\ %
	College of Science, Engineeering and Technolgy\\
	Texas Southern University\\
	Houston, TX 77004, USA\\
	\mbox{}\\ %
}

\date{}

\documentclass[12pt]{article}

\usepackage{graphicx}
\usepackage{subfigure} 
\usepackage{amsmath}
\usepackage{hyperref}
\usepackage[numbers,sort&compress]{natbib}
\usepackage{placeins}
\usepackage{stfloats}

\begin{document}
\maketitle

\begin{abstract}
Oil and gas pipeline leakages lead to not only enormous economic loss but also environmental disasters. How to detect the pipeline damages including leakages and cracks has attracted much research attention. One of the promising leakage detection method is to use lead zirconate titanate (PZT) transducers to detect the negative pressure wave when leakage occurs. PZT transducers can generate and detect guided stress waves for crack detection also. However, the negative pressure waves or guided stress waves may not be easily detected with environmental interference, e.g., the oil and gas pipelines in offshore environment. In this paper, a Gaussian mixture model based hidden Markov model (GMM-HMM) method is proposed to detect the pipeline leakage and crack depth in changing environment and time-varying operational conditions. Leakages in different sections or crack depths are considered as different states in hidden Markov models (HMM). Laboratory experiments show that the GMM-HMM method can recognize the crack depth and leakage of pipeline such as whether there is a leakage, where the leakage is.
\end{abstract}

\section{Introduction}
Thousands of miles of pipelines crisscrossed on the Gulf of Mexico seafloor are the veins for offshore oil and gas industry of the U.S. or even the whole world, while the leaks and ruptures of those pipelines lead to not only enormous economic loss but also environmental disasters \cite{Ho2020}. In the last few decades, many pipeline structural health monitoring techniques have been used to monitor damages \cite{Ho2020, Adegboye2019}. One of the promising methods for pipeline leakage detection is based on the negative pressure wave (NPW) \cite{Zhang2019, Zhu2017, Hou2013, Liang2013, Li2009}. NPW is generated at the leak point when the fluid or gas escapes in the form of a high velocity jet \cite{Wikipedia2020}. Then the NPW propagates along pipeline in both directions, i.e., the upstream and downstream of leakage point. The NPW can be detected by Lead zirconate titanate (PZT) transducers. PZT transducers are made of piezoelectric materials which can convert mechanical energy to electrical energy and vice versa. This piezoelectric effect leads to PZT transducers work as passive sensors or active actuators. PZT transducers can be effectively used as passive sensors to catch the acoustic signals propagating along the pipeline. On the other hand, the pipeline health condition needs to be evaluated periodically to provide early warning. To address this demand, PZT transducers have been used in active sensing mode to detect crack \cite{Li2019}. PZT transducers have found lots of applications for structural health monitoring. Wang et al. \cite{Wang2018} invented a wearable PZT transducers which can be easily and noninvasively “worn” onto the flanged valve for bolted joint in real-time. Gong et al. \cite{Gong2020} developed an algorithm to process signals collected by PZT sensor for automatic extraction of the stress wave reflection period. In these applications, the structural health condition is evaluated according to extracted features. However, in the actual situation, the changing environment and time-varying operational conditions make the reliability of damage evaluation facing the challenge. Especially for the offshore pipelines, the submarine environment is more complicated. With the development of machine learning, many attention has been paid to the probabilistic and statistical model-based methods, which are effective tools for characterizing uncertainties of signals such as Gaussian mixture model (GMM) \cite{Chakraborty2015}, hidden Markov model (HMM) \cite{Yang2018}. These researches have shown the potential to improve the damage evaluation reliability under uncertainties.  

Among the existing probabilistic and statistical models \cite{Chakraborty2015, Yang2018}, HMM has a strong capability in pattern classification, especially for signals with non-stationary natures and poor repeatability and reproducibility \cite{Li2007}. HMM and its variants have been extensively used for speech recognition \cite{Rabiner1989}, hand gesture recognition \cite{Wilson1999}, handwritten word recognition  \cite{Mohamed2000}, and newly applied to spam SMS detection \cite{Xia2020}. In the field of structural health monitoring, several researchers have also applied the HMM to damage evaluation. Rammohan and Taha \cite{Rammohan2005} using a standard HMM to model the simulated data of a pre-stressed concrete bridge. Tschöpe and Wolff \cite{Tschope2009} studied the HMM for damage degree classification on plate-like structures. These studies indicate that the HMM is robust to uncertainties. So far, however, there is very little published research on using HMM detect pipeline leakage.

\section{GMM-HMM based leakage detection}
In this section, method for damage indexes extraction from the acoustic signals detected by PZT sensors is discussed. Then the design and training of GMM-HMM are presented.
\label{sec:1}
\subsection{Damage indexes extraction}
To establish the relationship between characteristics change of the sampled waveform and damage parameters, two damage indexes are adopted to indicate the signal variations and serve as observations of the HMM model. The first damage index ($DI_{1}$) is a time-domain damage index, defined in \cite{Torkamani2014}:
\begin{equation} \label{eq1}
	DI_{1}=1-\sqrt{\frac{\int (s_1(t)-\bar{s}_1)(s_2(t)-\bar{s}_2)\,dt}{\int (s_1(t)-\bar{s}_1)^2\,dt \int (s_2(t)-\bar{s}_2)^2\,dt}},
\end{equation}
where $s_{1}(t)$ is the baseline waveform and $s_{2}(t)$ is the comparison waveform at time $t$. The $\bar{s}_1$ and  $\bar{s}_2$ is average value of $s_{1}(t)$ and $s_{2}(t)$. The baseline waveform represent the incident waveforms, and the comparison waveforms denote the captured waves detected by sensors. Unity minus the absolute value of the Pearson correlation coefficient is used as the time-domain damage index which can identify the signal difference.

The second one is a frequency-domain damage index ($DI_{2}$), the amplitude of peak frequency, as defined in \cite{Mei2016}:
\begin{equation} \label{eq2}
	DI_{2}=\max_{f_1\le f \le f_2}(|X(f)| ) ,
\end{equation}
where $X(f)=\int_{t_1}^{t_2} X(t) e^{-j2\pi ft} \,dt$. $f_{1}$ and $f_{2}$ are the start and stop frequency corresponding to the selected frequency spectrum window.

\label{sec:2}
\subsection{HMM and the Baum–Welch algorithm}
To reduce the calculation complexity and keep the model works efficiently, only first-order HMM is adopted in this study. A typical HMM model can be defined by a three-tuple:
\begin{equation} \label{eq3}
	\boldsymbol{\lambda}=\{\boldsymbol{\pi}, \boldsymbol{A}, \boldsymbol{B}\},
\end{equation}
where $\boldsymbol{\pi}$ is the initial probability distribution, $\boldsymbol{A}$ is the state transition probability matrix and $\boldsymbol{B}$ is the observation probability distribution matrix or the emission matrix. Common notations are used in this paper as follows.

a)	$\boldsymbol{S}$: A set of $N$ hidden states is denoted as $\boldsymbol{S} = \{s_1, s_2, s_3, … , s_N\}$. The state of model at time $t$ is denoted by $q_t \in \boldsymbol{S}$, which denotes the current state.

b)	$\boldsymbol{V}$: A set of $M$ observation states is denoted as $\boldsymbol{V} = \{v_1, v_2, v_3, … , v_M\}$. The observed state at time $t$ is denoted by $o_t \in \boldsymbol{V}$.

c)	$\boldsymbol{\pi}$: an $N\times1$ initial probability distribution over the state. $\pi_i$ is the probability that the Markov chain will start in the state $s_i$. Some states $s_j$ may have $\pi_j = 0$, meaning that they cannot be the initial state. Also, $\sum_{i=1}^{N} \pi_i=1$. 
\begin{equation} \label{eq4}
	\boldsymbol{\pi}=\{\pi_1, \pi_2, … ,\pi_N\}.
\end{equation}

d) $\boldsymbol{A}$: $N \times N$ state transition probability matrix, $N$ is the number of hidden states. $\boldsymbol{A}=[a_{ij}], 1\le i,j \le N$.
\begin{equation} \label{eq5}
	a_{ij}=P(q_{t+1}=s_j|q_t=s_i),
\end{equation}
where $a_{ij}$ is the probability of transmission from structural damage state $s_i$ at time $t$ to structural damage state $s_j$ at time $t+1$ and $\sum_{j=1}^{N} a_{ij}=1$.

e) $\boldsymbol{B}$: observation probability distribution matrix,
\begin{equation} \label{eq6}
	b_{jk}=P(o_t=v_k|q_t=s_j),
\end{equation}
where $ 1\le j \le N, 1\le k \le m$, $b_{jk}$ is the possibility of observation $o_t$ at time $t$. 

There are three basic problems in HMM: evaluation, decoding, and learning. Given a training set of observations, HMM is trained to find the optimized parameters $\boldsymbol{\lambda}=\{\boldsymbol{\pi}, \boldsymbol{A}, \boldsymbol{B}\}$, by Baum-Welch expectation-maximization algorithm (EM algorithm), which is a learning problem of HMM. After the HMM trained, this model can be applied to evaluate the damage states of a new defective pipeline in various conditional environments.

\begin{figure}[h]
	\centering
	\includegraphics[width= 10cm]{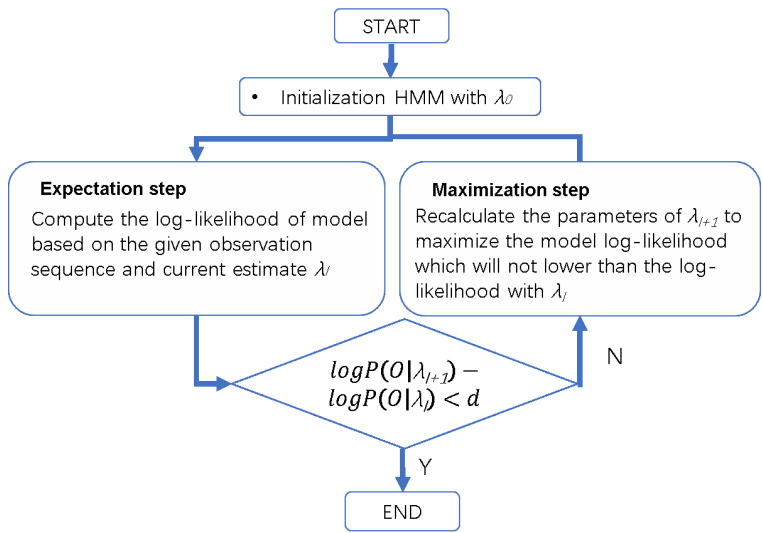}
	\caption{Baum-Welch algorithm (EM algorithm)}
	\label{fig1} 
\end{figure}

The Baum–Welch algorithm is employed in HMM to perform the training as shown in Fig. \ref{fig1}, which is an iterative process to adjust the parameters of $\boldsymbol{A}$, $\boldsymbol{B}$, and $\boldsymbol{\pi}$.

\label{sec:3}
\subsection{Gaussian mixture model}
In this paper, the Gaussian mixture model\cite{Mei2016,bishop2006pattern}, based on unsupervised learning, is chosen to model the distribution of the observed damage indexes to obtain $\boldsymbol{B}$ in the hidden Markov model, whose elements are given by equation
\begin{equation} \label{eq9}
	b_{jk}=P(o_t=v_k|q_t=S_j)=\sum_{k=1}^{M} c_{jk}N(O_t, \mu_{jk}, \Sigma_{jz}),
\end{equation}
where $c_{jk}$, $\mu_{jk}$ and $\Sigma_{jz}$ are the mixture weight, mean vector, the covariance matrix of the K-component Gaussian mixture model.

Based on the Gaussian distribution:
\begin{equation} \label{eq10}
	N(O_t, \mu_{jk}, \Sigma_{jz})=\frac{1}{\sqrt{2\pi \sigma^2}} \exp(\frac{-(x-\mu)^2}{2\sigma^2}).
\end{equation}

The EM algorithm is implied in the GMM model, which contains initialization, E step, and M step.

a)	Initialization

Setting the number of mixture component $K$, for each component initialize it with $c_{jk}$, $\mu_{jk}$ and $\Sigma_{jz}$.

b)	E step

Calculate the posterior probability based on current $c_{jk}$, $\mu_{jk}$ and $\Sigma_{jz}$.
\begin{equation} \label{eq11}
	\gamma(z_{tk}) = \frac{c_{jk}N(O_t, \mu_{jk}, \Sigma_{jz})}{\sum_{k=1}^{M} c_{jk}N(O_t, \mu_{jk}, \Sigma_{jz})}.
\end{equation}
\begin{equation} \label{eq12}
	\mu_{jk} = \frac{1}{N_k}\sum_{t=1}^{N}\gamma(z_{tk})O_t.
\end{equation}
\begin{equation} \label{eq13}
	\Sigma_{jz} = \frac{1}{N_k} \sum_{t=1}^{N}\gamma(z_{tk})(O_t-\mu_{jk})(O_t-\mu_{jk})^T.
\end{equation}
where $N_k = \sum_{t=1}^{N}\gamma(z_{tk})$.

c)	M step

Calculate new $c_{jk}$, $\mu_{jk}$, $\Sigma_{jz}$ based on the $\gamma(z_{tk})$.

d)	Iteration these steps until the GMM model converges.

\label{sec:4}
\subsection{GMM-HMM method}
There are five major steps included in the damage detection process. 

1) States of the HMM are designed based on the placement of pipelines and sensors. 

2) Stress waves are obtained from the sensors based on different states of the HMM. Damage indexes are extracted from the stress wave and describe the various degree of damages. The GMM model calculates the parameter of each Gaussian model which formed matrix $\boldsymbol{B}$ in HMM.

3) A typical HMM model defined by a three-tuple: $\boldsymbol{\lambda}=\{\boldsymbol{\pi}, \boldsymbol{A}, \boldsymbol{B}\}$. To initialize the HMM, $\boldsymbol{\pi}$ and $\boldsymbol{A}$ is estimate by the prior probability. 

4) After the parameter initialization of the HMM, the training implemented by an iterative process through the Baum-Welch algorithm.  

5) Given the trained HMM and the new observations to find out the optimal state is the typical decoding problem in HMM.

\section{Experimental Results}
\subsection{Leakage detection}
\subsubsection{Experiment setup}
The purpose of the experiment is to detect pipeline leakage utilizing PZT sensors. The negative pressure wave is generated by leakage in the pipeline and propagates along the pipeline from the leakage point to both ends. 

\begin{figure}
	\centering
	\includegraphics[width=0.8\textwidth]{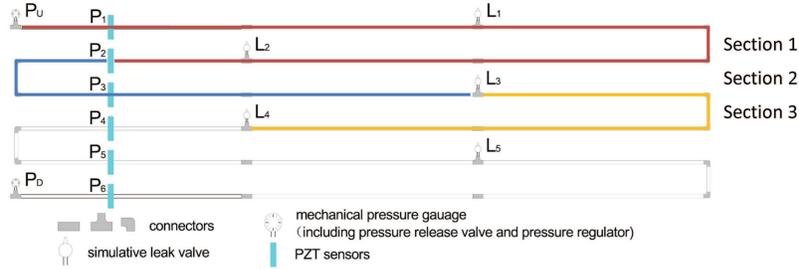}
	\caption{Settlement of the pipeline and sensors}
	\label{fig3}       
\end{figure}

The experimental pipeline built at the University of Houston, as shown in Fig. \ref{fig3}, consists of a series of PVC plain-end pipe sections connected together to form a pipeline with a total length of 55.78 $m$. Six PZT sensors ($P_1$ to $P_6$, size is 15 $mm \times 10$ $mm$) are directly mounted on the pipeline to detect NPW signal arrival. A NI PXI-5105 digitizer is used as a data acquisition system. The digitizer is triggered by the voltage signal of PZT No. 1 with the trigger level at $-.02 V$ and all the signals from six PZT sensors are recorded simultaneously at a sampling rate of 100 $KS/s$. Although it is not targeted to solve the problem of submarine pipeline leakage detection, the experimental data is used to validate the proposed GMM-HMM method. 

Two states, leaking or not leaking, are chosen as the states in a left to right HMM model, where the transitions only go from one state to itself or to a unique follower. Leakage signal collected by $P_1$ and baseline simulation signal of three different sections of leakage serves as original data of the model.

\subsubsection{Results of leakage detection}
\begin{figure}[htbp]
	\centering
	\subfigure[Different states signal]{
		\includegraphics[width=5.5cm]{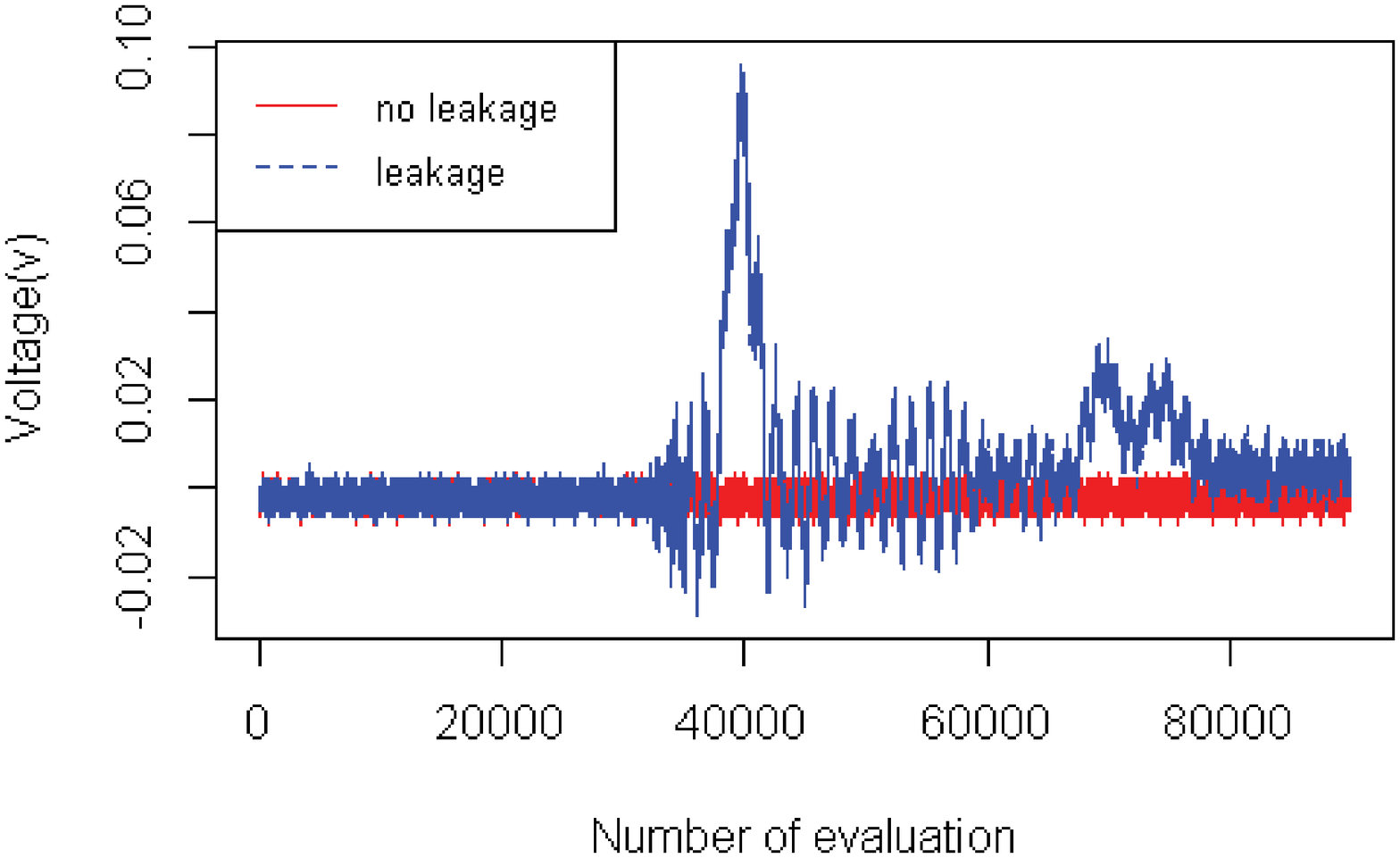}
	}
	\quad
	\subfigure[Damage indexes]{
		\includegraphics[width=5.5cm]{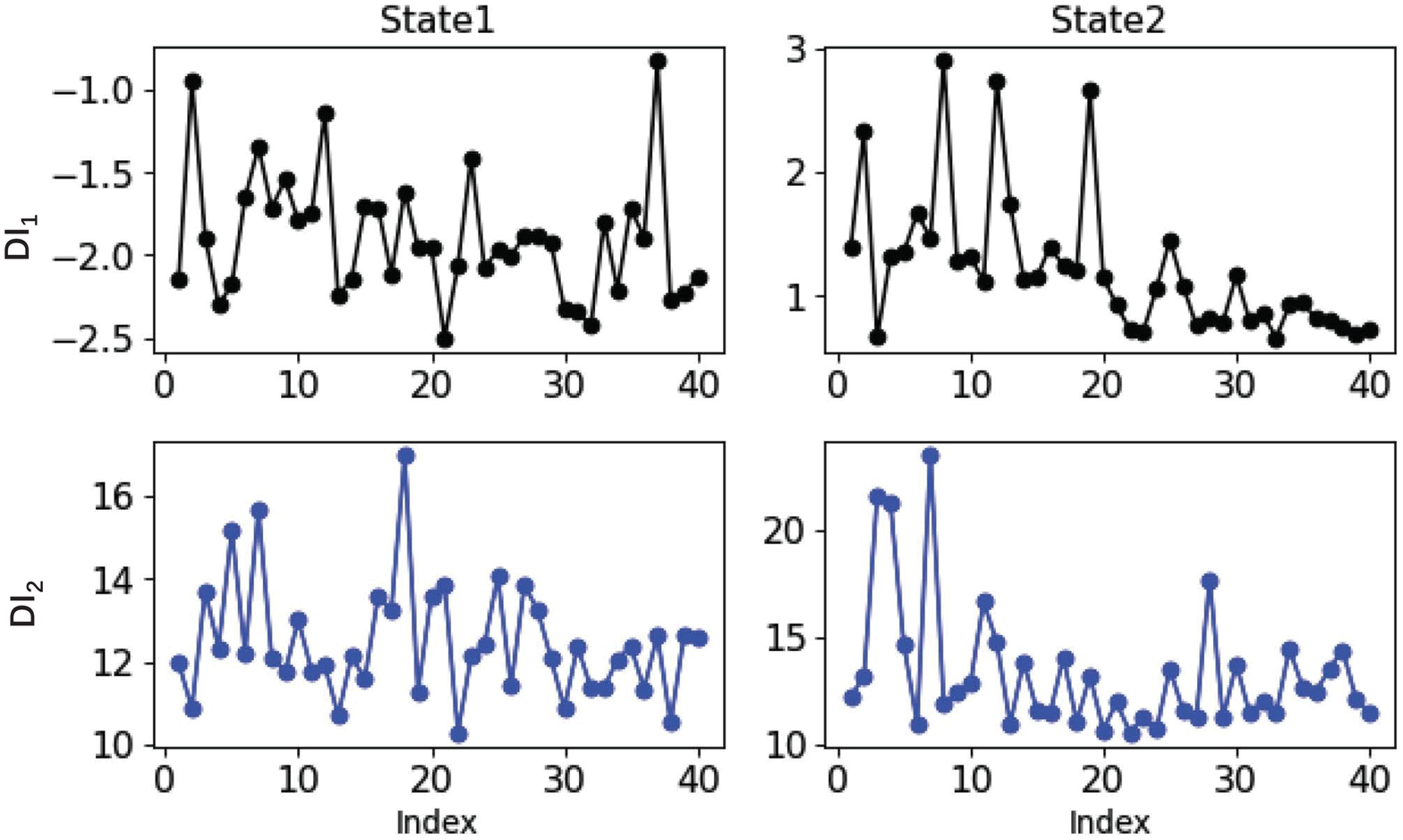}
	}
	\quad
	\subfigure[Negative log-likelihood]{
		\includegraphics[width=5.5cm]{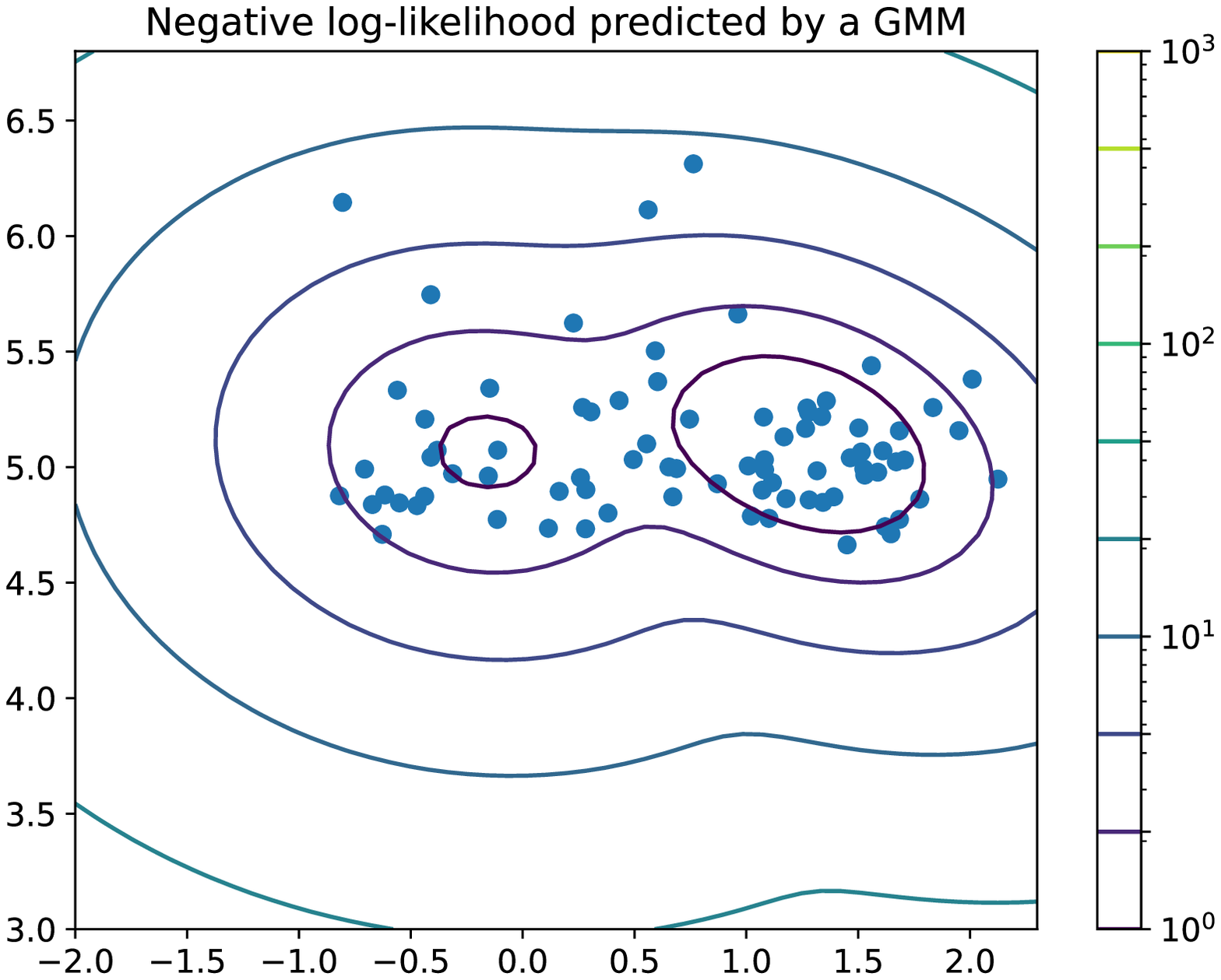}
	}
	\quad
	\subfigure[Pipeline status output]{
		\includegraphics[width=5.5cm]{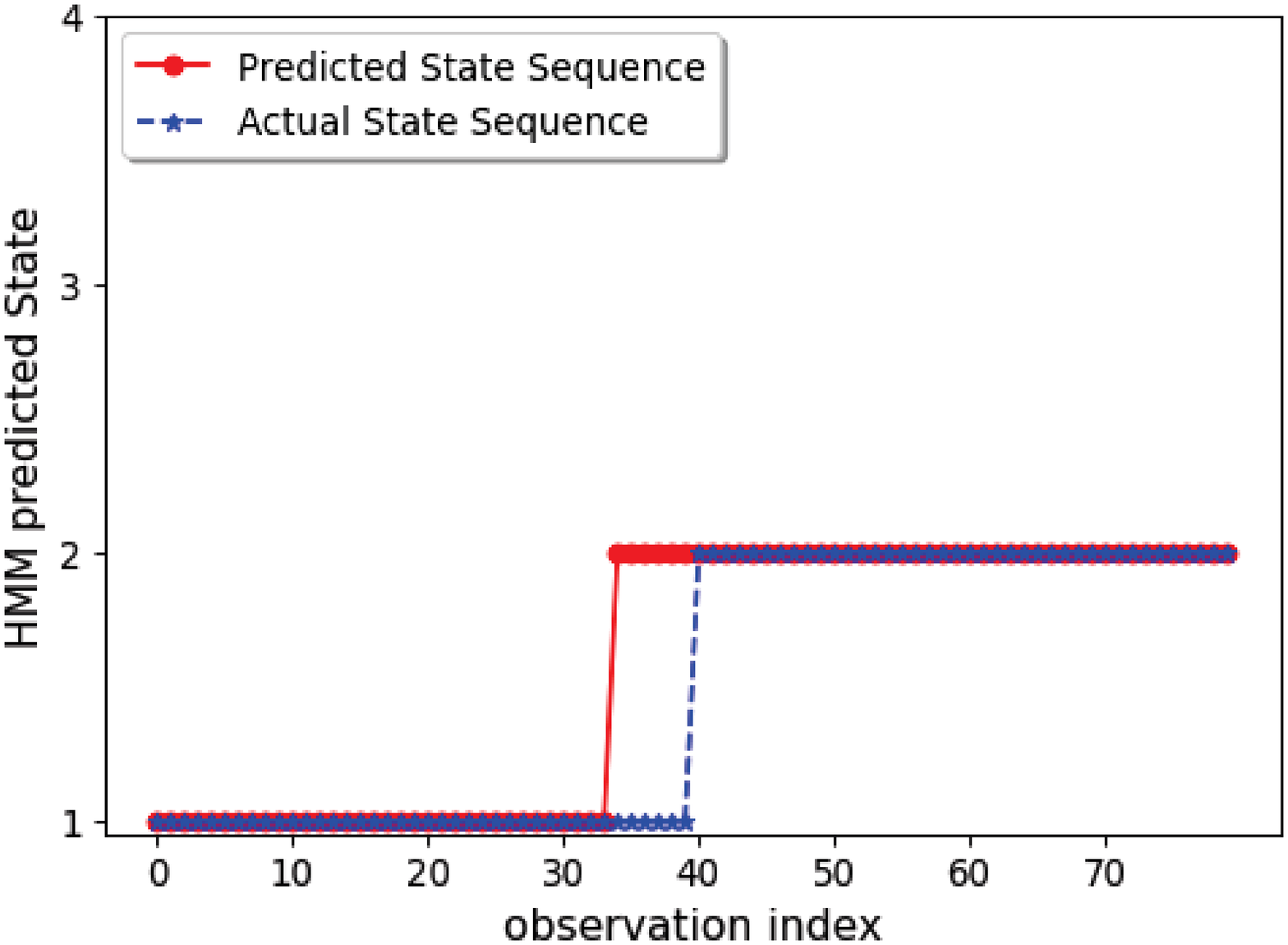}
	}
	\caption{Experimental results of leakage detection}
	\label{fig5}
\end{figure}

Signals of the different sections are shown in Fig. \ref{fig5}(a). The blue dashed line represents leakage state signal and red solid line represents no leakage state signal. The experiment of collecting signals for each leakage state was repeated by 20 times. Randomly selected repeated experiment data from the same sensor are separated into two groups equally, named training data and test data.

For training data, a total of 70 groups, 100000 per group samples are collected. Two damage indexes in different damage leakage states are obtained by using Eq. \ref{eq1} and Eq. \ref{eq2} with 45 data points for each state as shown in Fig. \ref{fig5}(b). The parameter of GMM can be calculated by EM algorithm.The negative log-likelihood predicted by GMM shown in Fig. \ref{fig5}(c) The testing is performed after the training process of the HMM model whose parameters have been optimized. The testing is used to validate the prediction capability of the HMM model and the result is showed in Fig. \ref{fig5}(d). The accuracy of the testing performance between predicted state sequence and actual state sequence are approaching 92.51\%.The accuracy is the ratio of correctly predicted states to the total number of state sequence.

\subsection{Leakage location detection}

\subsubsection{Experiment setup}
\begin{figure}[h!]
	\centering
	\subfigure[Schematic diagram of pipeline statues]{
		\includegraphics[width=6.5cm]{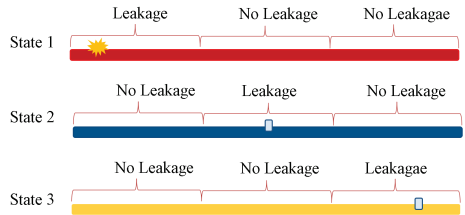}
	}

	\quad
	\subfigure[Signal of different states]{
		\includegraphics[width=5.5cm]{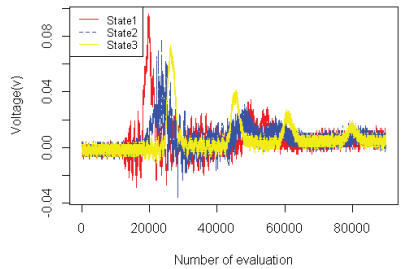}
	}

	\caption{Experimental results of leakage location detection}
	\label{fig6}
\end{figure}

In this experiment, all the data are based on the previous experimental setup. By changing the leakage state to different leakage locations. The purpose and output of the HMM model will be changed to leakage location detection. As shown in Fig. \ref{fig6}(a), there are three leakage locations corresponding to three different states in ergodic HMM. State 1 denotes leakage occurred at section 1 of pipeline, state 2 and 3 denotes leakage occurred at section 2 and 3 of pipeline. Therefore, there are three states in this ergodic HMM, which allowing for transitions from any emitting state to any other emitting state. 

Leakage signals are collected by $P_1$. As shown in Fig. \ref{fig6}(b), a total of 100 groups, 200000 per group samples are collected. To reduce the computational volume, data for this experience are cropped from the original data by 90000 data points to each state.

\subsubsection{Results of leakage location detection}
The 100 groups of measurements from three different leakage locations are separated into a training group and a test group. According to the observation density estimation method, the parameters of the Gaussian mixture model for each leakage state can be calculated by GMM based on EM algorithm. The maximum-likelihood number of Gaussian component was 3 estimated by GMM. The parameters of the HMM model are initialized as:
\begin{align*}
	\boldsymbol{\pi_0} = [1, 0, 0],
	\boldsymbol{A}={\left[ \begin{array}{cccc}
			1/3 &    1/3 &    1/3 	\\
			1/3 &    1/3 &    1/3   \\
			1/3 &    1/3 &    1/3
		\end{array} 
		\right ]}.
\end{align*}

The parameters of the three states HMM model are reassessed by the Baum-Welch algorithm. After training the updated parameters obtained are as follows:
\begin{align*}
	\boldsymbol{\pi} = [0, 0, 1],
	\boldsymbol{A}={\left[ \begin{array}{cccc}
			0.232 &    0.357 &    0.411 	\\
			0 	  &    0.525 &    0.475     \\
			0.539 &    0.102 &    0.359
		\end{array} 
		\right ]}.
\end{align*}

\begin{figure}[h!]
	\centering
	\includegraphics[width=0.7\textwidth]{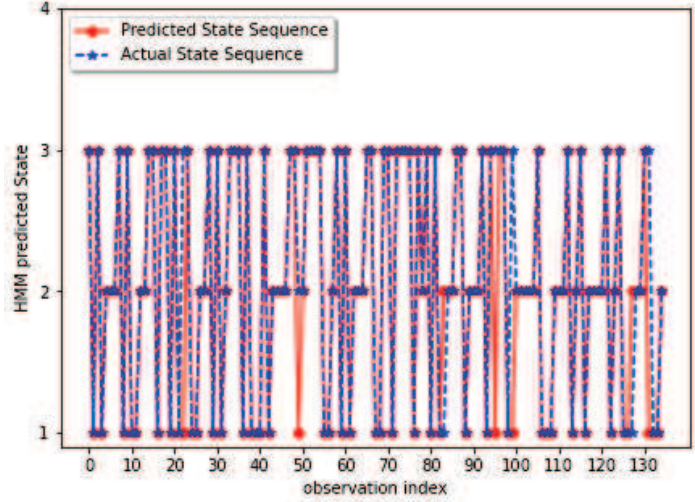}
	\caption{Actual state vs. HMM predict state}
	\label{fig7}       
\end{figure}
Then, the testing is performed after the parameter optimization of HMM model. The model calculates the posterior probability when new testing data are fed. By maximizing the posterior probability, the predicted state will be obtained. The leakage location evaluation result is shown in Fig. \ref{fig7}. The accuracy of the testing performance between predicted state sequence and actual state sequence are approaching 94.81\%.

\subsection{Crack depth inspection}

\subsubsection{Experiment setup}

\begin{figure}[h]
	\centering
	\includegraphics[width=0.9\textwidth]{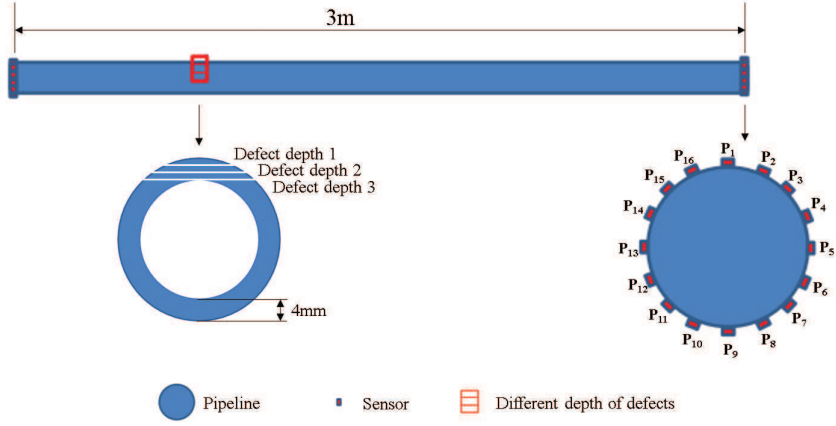}
	\caption{Setup of the defect inspection experiment}
	\label{fig8}       
\end{figure}

The purpose of the experiment is to detect the depth of cracks utilizing PZT sensors. The experimental pipeline built at the University of Houston, consists of a section of galvanized steel pipe with a total length of 3 meters. A PZT array with sixteen PZT transducers are directly mounted on the pipeline to detect defection signal.Shown in Fig. \ref{fig8}. The guided wave is generated from sensors at the left side of pipe at the frequency of 50 kHz, propagates along the pipeline through the defect point, and received by the sensors at other side. This PZT array can detect the pipeline defect location based on time-reversal method and matching pursuit de-noising \cite{Xu2019}. In addition, this PZT array has been used for underwater communication by using stress wave propagation along pipelines \cite{He2020}.

Three states, defect depth 1, defect depth 2, and defect depth 3, are chosen as the states in the left to right HMM model, which means the depth of the defect was increased unidirectionally. Received defect signal is collected from the right-hand side sensors. 

\subsubsection{Results of depth inspection}	
The 154 group measurements from three defect depth inspection are separated into a training group and a test group. The parameters of the Gaussian mixture model are also calculated based on the observation density estimation method, where the number of Gaussian is set to 3. The parameters of the left to right HMM model are initialized as:

\begin{align*}
	\boldsymbol{\pi_0} = [1, 0, 0],
	\boldsymbol{A}={\left[ \begin{array}{cccc}
			0.9 &    0.1 &    0 	\\
			0   &    0.9 &    0.1   \\
			0   &    0 &      1
		\end{array} 
		\right ]}.
\end{align*}

The parameters of the three states HMM model are also reassessed by the Baum-Welch algorithm. After training the updated parameters obtained are as follows:

\begin{align*}
	\boldsymbol{\pi} = [1, 0, 0],
	\boldsymbol{A}={\left[ \begin{array}{cccc}
			0.974 &    0.026 &    0 	\\
			0   &    0.976 &    0.024   \\
			0   &    0 &      1
		\end{array} 
		\right ]}.
\end{align*}

\begin{figure}[h]
	\centering
	\includegraphics[width=0.7\textwidth]{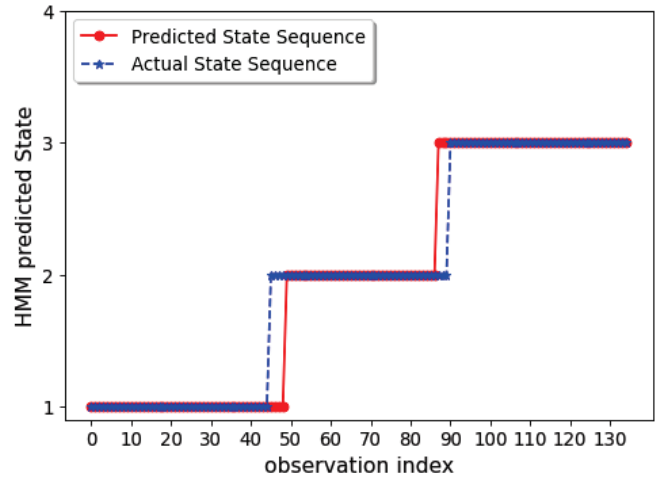}
	\caption{HMM output}
	\label{fig11}       
\end{figure}

By maximizing the posterior probability, the predicted state will be obtained. The crack depth inspection result is shown in Fig. \ref{fig11}. The accuracy of testing performance between predicted state sequence and actual state sequence are approaching 93.23\%.

\section{Conclusions}\label{conclusions}
Pipeline leakage detection and crack depth identification are difficult especially in changing environment and time-varying operational conditions. This study applied a GMM-HMM based on a probabilistic damage evaluation method on pipeline leakage detection. The GMM-HMM method calculates the maximum log-likelihood recursively, which reveals the emission probability from hidden states to observable states. Thus, HMM can have a better description of the leakage by using observations from GMM which contains implicit damage indexes. The experiment performed on the pipeline showed great potential to deal with various leakage detection problems. 

\paragraph{Funding}
This research was funded by National Science Foundation grant number 1801811.

\paragraph{Acknowledgments}
Thanks to Smart Materials and Structures Laboratory at the University of Houston for providing the pipeline leakage experimental data.

\bibliographystyle{IEEEtran}
\bibliography{bibfile}

\end{document}